\documentclass[useAMS,usenatbib]{mn2e}

\title[Low-Frequency Structure of the Crab Pulsar]
{Interpretation of the Low-Frequency Peculiarities in the Radio
Profile Structure of the Crab Pulsar}

\author[S. A. Petrova]{S. A. Petrova
\thanks{E-mail: petrova@ira.kharkov.ua}\\
Institute of Radio Astronomy, NAS of Ukraine, 4, Chervonopraporna
Str., 61002 Kharkov, Ukraine}

\begin{document}

\date{Received\dots}

\pagerange{\pageref{firstpage}--\pageref{lastpage}} \pubyear{2008}

\maketitle

\label{firstpage}

\begin{abstract}
The theory of magnetized induced scattering off relativistic
gyrating particles is developed. It is directly applicable to the
magnetosphere of a pulsar, in which case the particles acquire
gyration energies as a result of resonant absorption of radio
emission. In the course of the radio beam scattering into
background the scattered radiation concentrates along the ambient
magnetic field. The scattering from different harmonics of the
particle gyrofrequency takes place at different characteristic
altitudes in the magnetosphere and, because of the rotational
effect, gives rise to different components in the pulse profile.
It is demonstrated that the induced scattering from the first
harmonic into the state under the resonance can account for the
so-called low-frequency component in the radio profile of the Crab
pulsar. The precursor component is believed to result from the
induced scattering between the two states well below the
resonance. It is shown that these ideas are strongly supported by
the polarization data observed. Based on an analysis of the
fluctuation behaviour of the scattering efficiencies, the
transient components of a similar nature are predicted for other
pulsars.
\end{abstract}


\section{Introduction}
\subsection{Radio emission components outside of the main pulse}

The Crab pulsar is known for its complex radio profile
\citep[e.g.][]{mh96}. It is built of a total of seven components,
which are spread out over the whole pulse period and exhibit
substantially distinct spectral and polarization properties. At
the lowest frequencies, $\la 600$ MHz, the profile consists of the
three components: the main pulse (MP), the precursor (PR) $\sim
15^\circ$ ahead of the MP and the interpulse (IP), which lags the
MP by $\sim 150^\circ$ and is connected to it by a weak emission
bridge \citep*{r70,mht72,v73}. The PR component is distinguished
by its complete linear polarization and extremely steep spectrum.
At frequencies $\ga 1$ GHz, where the PR has already vanished,
there appears another component $\sim 36^\circ$ in advance of the
MP \citep{mh96}. This so-called low-frequency component (LFC) is
broader and weaker than the PR. Moreover, the percentage of linear
polarization in the LFC is less, $\sim 40\%$ \citep{mh98}, though
still markedly exceeds that in the MP and IP ($\sim 25\%$ and
$15\%$, respectively).

The IP and LFC become invisible at $\sim 3$ and 5 GHz, and at
still higher frequencies the profile structure changes drastically
\citep{mh96,mh98}. In addition to the MP, there is the interpulse
(IP'), which re-appears some $10^\circ$ earlier in phase, and two
high-frequency components (HFC1 and HFC2), $\sim 70^\circ$ and
$130^\circ$ behind the IP'. All these new components are
characterized by high linear polarization and relatively flat
spectra, so that at frequencies $\ga 8$ GHz the MP disappears. The
fluctuation properties of these components are also worth noting.
According to the recent high-frequency studies, all the components
show occasional giant pulses \citep{h05,slow05}, the temporal and
frequency structure of the giant MPs and IP's being essentially
different \citep{hc07}.

The components outside of the MP are present in other pulsars as
well. About $40\%$ of the millisecond pulsars and $2\%$ of the
normal ones are known to exhibit IPs \citep{k99}. Besides that, a
handful of pulsars have firmly established PRs. Note that the PRs
are met in the profiles with IPs. Thus, the structure of the Crab
profile at the lowest frequencies is similar to that in some other
pulsars (e.g., PSR B1055-52, \citealt{m76} and PSR B1822-09,
\citealt*{f81}). However, in these cases the profile structure is
preserved over a wide frequency range, though the component
spectra are also somewhat different. Such components of the Crab
profile as the LFC, HFC1 and HFC2 as well as the high-frequency
shift of the IP to earlier pulse longitudes are unique. It should
be noted, however, that some millisecond pulsars have even more
complex structure (e.g., PSR J0437-4715, \citealt{mj95}), but it
is not well studied and classified.

The mechanisms of the IP and PR emissions are still a matter of
debate, while the nature of other components of the Crab profile
is completely obscure. The IP components are usually interpreted
in terms of geometrical models. It is assumed that the IP emission
originates in a distinct region (e.g. in the outer magnetosphere
or at the opposite magnetic pole) and can only be observable due
to a specific geometry of the pulsar (in cases of approximate
alignment or approximate orthogonality of the rotational and
magnetic axes). Recently \citet*{d05} have developed a generalized
geometrical model for the PSR B1822-09, which includes the
formation of the PR as well. It has been suggested that the PR
component originates well above the MP and the backward emission
from this region forms the IP. Such a pattern can be observable if
the pulsar is a nearly orthogonal rotator.

It should be noted that the geometrical models are insufficient to
account for the bulk of observational facts. Firstly, the
components outside of the MP show peculiar polarization and
spectral properties. Secondly, the fluctuation behaviour of the
components strongly testifies to their physical connection with
the MP. In particular, the subpulse modulation in the MP and IP of
the PSR B1702-19 has been found to be intrinsically in phase
\citep*{welt07}. All this calls for a physical interpretation.

Recently we have proposed a physical mechanism of the PR and IP
components based on propagation effects in pulsar magnetosphere
\citep{p07a,p07b}. These components are suggested to result from
induced scattering of the MP emission into the background. In case
of efficient scattering, the scattered radiation grows
substantially and concentrates in the direction corresponding to
the maximum scattering probability. In the regime of a superstrong
magnetic field, the scattered component is directed along the
ambient field and can be identified with the PR. In a moderately
strong magnetic field, the radiation is predominantly scattered in
the opposite direction, giving rise to the IP. Within the
framework of this model, the basic features of the components as
well as their connection to the MP are explained naturally. Our
theory can be elaborated further to explain the complicated radio
emission pattern of the Crab pulsar. The present paper is devoted
to the generalized mechanism of the two components, the PR and
LFC, which precede the MP and develop at relatively low
frequencies. The formation of the high-frequency components of the
Crab will be addressed in our forthcoming paper. It will be argued
that the IP' results from the backward scattering of the PR, while
the HFC1 and HFC2 present the backscattered emission of the LFC.

\subsection{Statement of the problem}
The magnetosphere of a pulsar contains the ultrarelativistic
electron-positron plasma, which streams outwards along the open
magnetic field lines and leaves the magnetosphere as a pulsar
wind. The pulsar radio emission is generally believed to originate
deep in the open field line tube, and on its way in the
magnetosphere it should propagate through the plasma flow. As the
brightness temperatures of the pulsar radio emission are extremely
high, one can expect that induced scattering off the plasma
particles is significant. Deep in the magnetosphere the magnetic
field is strong enough to affect the scattering process
considerably by modifying both the scattering cross-section and
the particle recoil. This happens for the waves below the
cyclotron resonance, as long as the frequency in the particle rest
frame remains much less than the electron gyrofrequency,
$\omega^\prime\ll\omega_G\equiv eB/mc$. The magnetized induced
scattering in pulsars is known to be efficient \citep{bs76} and is
suggested to have a number of observational consequences
\citep{lp96,p04a,p04b,p07a,p07b}. As the magnetic field strength
decreases with distance from the neutron star, in the outer
magnetosphere the radio waves pass through the resonance. The
scattering by the pulsar wind holds in the non-magnetic regime and
can also be efficient in pulsars \citep{wr78,lp96}. Note that in
the resonance region itself the radio waves are subject to
resonant absorption rather than scattering
\citep*{bs76,lp98,p02,p03,melr_a,melr_b}.

Close to the neutron star surface, the magnetic field is so strong
that any perpendicular momentum of the particles is almost
immediately lost via synchrotron re-emission. Hence, the particles
are confined to the magnetic field lines, and it is usually
assumed that they perform ultrarelativistic rectilinear motion
throughout the open field line tube. However, in the outer
magnetosphere, where synchrotron re-emission is already
inefficient, the particles can easily gain relativistic gyration
energies as a result of resonant absorption of the radio emission
\citep{lp98,p02,p03}. As has been shown in \citet{p02,p03}, the
absorbing particles reach relativistic gyration at the very bottom
of the resonance region for radio waves, in the course of
absorption of the highest-frequency waves, $\nu\ga 10$ GHz. Then
the lower-frequency waves, $\nu\ll 10$ GHz, which are still below
the resonance, $\omega^\prime\ll \omega_G$, are scattered by the
particles performing relativistic helical motion.

The spontaneous scattering off a gyrating electron in application
to pulsars has recently been studied in \citet{p07c}. The process
has been found to differ substantially from the scattering by the
particle at rest. The total cross-section of the former process
appears much larger, and the scattered waves predominantly
concentrate at high harmonics of the gyrofrequency, close to the
maximum of the synchrotron emission of the scattering electron,
whereas the scattering by the particle at rest leaves the wave
frequency unchanged. Although the spontaneous scattering by
gyrating particles appears weak enough to markedly suppress pulsar
radiation, except for the lowest frequencies, $\nu <100$ MHz, it
supplements the synchrotron re-emission in reprocessing the radio
photons into the high-energy emission of the optical or soft X-ray
band.

All the previous studies of induced scattering in pulsars have
assumed rectilinear motion of the scattering particles. It has
been demonstrated that even in the regime of a superstrong
magnetic field the induced scattering is most efficient well above
the emission region, at distances roughly comparable with the
radius of cyclotron resonance \citep{p07a}. Therefore the gyration
of the scattering particles needs to be taken into account. In the
present paper, we examine the induced scattering off the spiraling
particles. In contrast to the spontaneous scattering, the induced
scattering between the states corresponding to close harmonics of
the gyrofrequency is more efficient. We consider the induced
scattering between the zeroth-harmonic states, when the incident
and final frequencies are both below the resonance, study the
scattering from several first harmonics to the zeroth one and
conclude that these processes can be efficient in pulsars. The
kinetic equations derived are applied to the problem of the radio
beam scattering into the background. It is found that the
radiation is predominantly scattered along the ambient magnetic
field. Since the scattering regions lie at different altitudes
above the emission region, the rotational aberration places the
scattered components at different pulse longitudes ahead of the
MP. Proceeding from the numerical estimate of the radius of
cyclotron resonance in the Crab pulsar we show that the PR
component should result from the scattering under the resonance
and the LFC from the first-harmonic scattering.

The plan of the paper is as follows. Basic formalism of induced
scattering off the spiraling particles is developed in Sect.~2.
The problem on the radio beam scattering into the background is
examined in Sect.~3. Applications to the the Crab pulsar are
presented in Sect.~4. Our results are discussed in Sect.~5 and
briefly summarized in Sect.~6.

\section{Basic formalism of induced scattering}
Let us consider the induced scattering of transverse
electromagnetic waves by the system of particles in the presence
of an external magnetic field. The scattering particles are
assumed to perform relativistic helical motion. The rate of change
of the photon occupation number $N$ as a result of the scattering
is given by
\begin{equation}
\frac{\partial N}{\partial t}=n_e\int wNN_1\frac{{\rm d}^3{\bmath
k_1 }}{(2\pi)^3}\frac{\partial f}{\partial{\bmath p}}\Delta{\bmath
p}{\rm d^3}{\bmath p},
\end{equation}
where $N=N(\bmath k)$, $N_1=N_1(\bmath k_1)$, $\bmath k$ and
$\bmath k_1$ are the photon wavevectors in the initial and final
states, $w=w({\bmath k},{\bmath k_1},{\bmath p})$ is the
scattering probability, $n_e$ is the number density of the
scattering particles, $f(\bmath p)$ is the particle distribution
function in momenta normalized as $\int f(\bmath p){\rm
d}^3{\bmath p}=1$, $\Delta\bmath p$ is the momentum increment in
the scattering act. In case of the scattering by the particles at
rest, the recoil is extremely small, $\Delta k/k\sim \hbar k/mc<<<
1$, so that the photon frequency is approximately unaltered in the
scattering act, $\omega_1^{(\rm r)}\approx \omega^{(\rm r)}$. In
case of the scattering by the spiraling particles, the situation
is essentially distinct: The frequencies of the scattered photons
present a discrete set. In the guiding centre frame,
\begin{equation}
\omega_1^{(\rm c)}=\omega^{(\rm c)}+s\omega_H/\gamma_0,\quad
s=0,\pm 1,\pm 2,\dots ,
\end{equation}
where $\gamma_0$ is the Lorentz-factor of the circular motion of
the particle, $\gamma_0\equiv (1-\beta_0^2)^{-1/2}$, $\beta_0$ is
the particle velocity in units of $c$. In the laboratory frame,
equation (2) is written as
\begin{equation}
\omega_1\gamma_\Vert\eta_1=\omega\gamma_\Vert\eta+s\omega_H/\gamma_0,\quad
s=0,\pm 1,\pm 2,\dots ,
\end{equation}
where $\eta=1-\beta_\Vert\cos\theta$,
$\eta_1=1-\beta_\Vert\cos\theta_1$, $\beta_\Vert$ is the
longitudinal component of the particle velocity normalized to $c$,
$\theta$ and $\theta_1$ are the tilts of the initial and final
wavevectors to the magnetic field,
$\gamma_\Vert\equiv(1-\beta_\Vert^2)^{-1/2}$.

For a given harmonic of the gyrofrequency, $s$, the scattering
probability is related to the differential cross-section per
elementary solid angle ${\rm d}O_1$ as
\begin{equation}
w_s=(2\pi)^3c^4\frac{\omega}{\omega_1^3}\frac{{\rm
d}\sigma_s}{{\rm
d}O_1}\delta(\omega_1-\omega\eta/\eta_1-s\omega_H/\gamma\eta_1)
\end{equation}
and the relativistic transformation of the cross-section from the
guiding-centre frame reads
\begin{equation}
\frac{{\rm d}\sigma}{{\rm d}O_1}=\left (\frac{{\rm d}\sigma}{{\rm
d }O_1}\right )^{({\rm c})}\frac{\eta^2}{\gamma_\Vert^2\eta_1^3}
\end{equation}
\citep[for more details see][]{p07c}. In equation (4) above,
$\gamma$ stands for the total Lorentz-factor of the particle, and
using the invariance of the transverse momentum,
$p_\perp=\beta_\perp\gamma mc=\beta_0\gamma_0mc$, it is easy to
see that $\gamma=\gamma_0\gamma_\Vert$. A general form of the
scattering cross-section in the guiding centre frame is given by
eqs. (11)-(12) in \citet{p07c}. It actually includes four
cross-sections, which correspond to one of the two possible
polarizations of the photons in the initial and final states. The
photons are polarized either in the plane of the wavevector and
the ambient magnetic field (A-polarization) or perpendicularly to
this plane (B-polarization).

Given that $\beta_\Vert=0$, $s=0$ and $\omega=\omega_1\ll
\omega_H/\gamma_0$, the cross-sections are reduced to \citep[see
also eq. (17) in][]{p07c}
\[ \frac{{\rm d}\sigma_0^{AA}}{{\rm
d}O_1}=\frac{r_e^2\sin^2\theta\sin^2\theta_1}{\gamma_0^2},
\]
\[ \frac{{\rm d}\sigma_0^{AB}}{{\rm
d}O_1}=\frac{r_e^2\omega^2}{\omega_H^2}\left
[\cos\theta\cos\Delta\phi -\frac{\beta_0^2\sin\theta
\sin\theta_1\cos\theta_1}{2}\right ]^2, \]
\[ \frac{{\rm d}\sigma_0^{BA}}{{\rm
d}O_1}=\frac{r_e^2\omega^2}{\omega_H^2}\left
[\cos\theta_1\cos\Delta\phi -\frac{\beta_0^2\sin\theta
\sin\theta_1\cos\theta}{2}\right ]^2, \]
\begin{equation}
\frac{{\rm d}\sigma_0^{BB}}{{\rm
d}O_1}=\frac{r_e^2\omega^2}{\omega_H^2}\sin^2\Delta\phi,
\end{equation}
where $\Delta\phi$ is the difference of the azimuthal coordinates
of the wavevectors in the initial and final states. Besides that,
we are interested in the induced scattering between the states,
one of which still corresponds to the low frequency,
$\omega_1\ll\omega_H/\gamma_0$, while in another one
$\omega\approx s\omega_H/\gamma_0$. The induced scattering is
believed to be efficient for small enough $s$, $s\sim 1$, and the
scattering between the states with $\omega\sim n\omega_H/\gamma_0$
and $\omega_1\sim l\omega_H/\gamma_0$, where $n,l\gg 1$, should be
much weaker. In the case of interest, the cross-sections take the
form
\[ \frac{{\rm d}\sigma_s^{AA}}{{\rm
d}O_1}=\frac{r_e^2}{\gamma_0^2}J_s^2(s\beta_0\sin\theta)
\sin^2\theta\sin^2\theta_1\cot^4\theta,\]
\[ \frac{{\rm d}\sigma_s^{AB}}{{\rm
d}O_1}=\frac{r_e^2\omega_1^2}{\omega_H^2}J_s^2(s\beta\sin\theta)\cot^4\theta\]
\[\times\left\{\sin\theta\cos\Delta\phi+\sin\theta_1\left [
1-\beta_0^2(1-\cos\theta\cos\theta_1)/2\right ]\right\}^2,\]
\[ \frac{{\rm d}\sigma_s^{BA}}{{\rm
d}O_1}=\frac{r_e^2}{\gamma_0^2}J_s^{\prime^2}(s\beta_0\sin\theta)
\beta_0^2\cos^2\theta\sin^2\theta_1,\]
\[ \frac{{\rm d}\sigma_s^{BB}}{{\rm
d}O_1}=\frac{r_e^2\beta_0^2\omega_1^2}{\omega_H^2}J_s^{\prime^2}(s\beta\sin\theta)\]
\begin{equation}
\times\left\{\sin\theta\cos\Delta\phi+\sin\theta_1\left [
1-\beta_0^2(1-\cos\theta\cos\theta_1)/2\right ]\right\}^2.
\end{equation}
For the inverse scattering, $\omega\ll\omega_H/\gamma_0$,
$\omega_1\approx s\omega_H/\gamma_0$, the cross-sections are given
by eq. (18) in \citet{p07c}. Comparing those results with equation
(7) above, one can see that the scattering probability (4) is
symmetrical with respect to the initial and final photon states.
This is known to be the fundamental property of this quantity.

The momentum increment $\Delta\bmath p$ can be found from the
conservation laws in the elementary scattering act. In the
presence of an external magnetic field, the energy and the
momentum component parallel to the field are conserved,
\[\Delta\gamma mc^2=\hbar (\omega-\omega_1),\]
\begin{equation}
\Delta p_\Vert=\hbar (k\cos\theta-k_1\cos\theta_1).
\end{equation}
Using equation (3) and differentiating the relation
$\gamma^2m^2c^4\equiv m^2c^4+(p_\perp^2+p_\Vert^2)c^2$, one can
find that
\begin{equation}
\Delta p_\perp =-s\hbar\omega_H\frac{m}{p_\perp}.
\end{equation}
Thus, the perpendicular momentum is changed if $s\neq 0$.

\section{Induced scattering of pulsar radio beam into background}
Now let us concentrate on the problem of induced scattering into
background based on the general theory presented above. Pulsar
radiation is known to be highly directional: At any point of the
emission cone it is concentrated into a beam of the opening angle
$\la 1/\gamma$, which is typically much less than the angular
width of the emission cone. Therefore at any point of the
scattering region the incident radiation can be approximately
presented by a single wavevector $\bmath k$. Far enough from the
emission region, the radiation propagates quasi-transversely with
respect to the ambient magnetic field, $1/\gamma\ll\theta\la 1$.

Since the rate of induced scattering depends on the photon
occupation number in the final state, the scattering out of the
beam is impossible unless there are some background photons which
trigger the scattering process. Such photons should indeed be
present. In particular, they may result from the spontaneous
scattering from the radio beam. Although the occupation numbers of
the background are extremely small, they may still stimulate
efficient induced scattering from the beam. Then the beam photons
are predominantly scattered into the state $\bmath k_1$
corresponding to the maximum scattering probability, and the
photon occupation number in this state may become comparable with
the original occupation number of the beam photons. Thus, we come
to the problem on induced scattering between the two photon
states, $\bmath k$ and $\bmath k_1$; the parameters of the
background state corresponding to the maximum scattering
probability will be specified below based on an analysis of the
concrete kinetic equations.

\subsection{The case $s=0$}
We start from the case when the frequency in the guiding centre
frame remains unchanged in the scattering act, i.e. $s=0$ and
$\omega\eta=\omega_1\eta_1$. Then $\Delta p_\perp =0$ and the
scattering is analogous to that off the rectilinearly moving
particles. The latter has been examined in \citet{p07a}, and now
we are interested if there are any quantitative differences
between the two processes. It is convenient to replace the photon
occupation numbers $N$ and $N_1$ with the intensities $i_\nu$ and
$i_{\nu_1}$ defined as
\[i_\nu (\nu,\theta,\phi){\rm d}\nu{\rm d}O\equiv
2\hbar\omega cN({\bmath k})\frac{{\rm d}^3{\bmath k}}{(2\pi)^3},\]
\begin{equation}
i_{\nu_1} (\nu_1,\theta_1,\phi_1){\rm d}\nu_1{\rm d}O_1\equiv
2\hbar\omega_1 cN({\bmath k_1})\frac{{\rm d}^3{\bmath
k_1}}{(2\pi)^3},
\end{equation}
and to perform trivial integration over the solid angle,
introducing the spectral intensities $I_\nu\equiv\int i_\nu{\rm
d}O$ and $I_{\nu_1}\equiv\int i_{\nu_1}{\rm d}O_1$. The scattering
is assumed to be stationary: The intensity varies only on account
of the beam propagation through the scattering region, whereas the
parameters of the scattering process do not depend on time, and
therefore $\partial /\partial t=c{\rm d}/{\rm d}r$.

Recall that the scattering cross-sections for gyrating particles
given by equation (6) are obtained in the approximation
$\omega^{(\rm c)}\ll\omega_H/\gamma_0$. One can see that $\left
({\rm d }\sigma^{AB}_0/{\rm d}O_1,{\rm d }\sigma^{BA}_0/{\rm
d}O_1,{\rm d }\sigma^{BB}_0/{\rm d}O_1\right )/\left ({\rm d
}\sigma^{AA}_0/{\rm d}O_1\right )
\sim\omega^{(c)^2}\gamma_0^2/\omega_H^2\ll 1$, i.e. the scattering
in the polarization channel $A\to A$ strongly dominates those in
the other channels. Note that the same polarization signature is
also characteristic of the scattering by the particle at rest.

The cross-section ${\rm d }\sigma^{AA}_0/{\rm d}O_1$ can be
rewritten in terms of the quantities of the laboratory frame using
the relativistic transformations $\sin\theta^{(\rm
c)}=\sin\theta/\gamma_\Vert\eta$ and $\sin\theta_1^{(\rm
c)}=\sin\theta_1/\gamma_\Vert\eta_1$. Then, substituting equations
(4) and (5) into equation (1) and performing integration over the
wavenumber with the help of delta-function, one can obtain
\[
\frac{{\rm d}I_\nu}{{\rm d}r}=I_\nu
I_{\nu_1}r_e^2n_ec\int\frac{\cos\theta
-\cos\theta_1}{2\nu_1^2\eta_1^2}\frac{\sin^2\theta\sin^2\theta_1}
{\eta^2\eta_1^2}\]
\begin{equation}
\times\frac{m^2c^2p_\perp^4}{p_\Vert^6}\frac{\partial f }{\partial
p_\Vert}{\rm d}p_\Vert p_\perp{\rm d}p_\perp .
\end{equation}
Here it is taken into account that $\gamma_\Vert=\gamma/\gamma_0$,
$p_\Vert =\beta_\Vert\gamma mc\approx \gamma mc$ and $p_\perp
=\beta_0\gamma_0mc\approx\gamma_0mc$. For the inverse scattering
we find analogously
\[
\frac{{\rm d}I_{\nu_1}}{{\rm d}r}=I_\nu
I_{\nu_1}r_e^2n_ec\int\frac{\cos\theta_1
-\cos\theta}{2\nu^2\eta^2}\frac{\sin^2\theta\sin^2\theta_1}
{\eta^2\eta_1^2}\]
\begin{equation}
\times\frac{m^2c^2p_\perp^4}{p_\Vert^6}\frac{\partial f }{\partial
p_\Vert}{\rm d}p_\Vert p_\perp{\rm d}p_\perp .
\end{equation}
Keeping in mind that $\nu\eta=\nu_1\eta_1$, one can see that the
right-hand sides of equations (11) and (12) are equal in absolute
value and have opposite signs. Performing integration by parts and
noticing that the scattering probability peaks at $\theta_1^{\rm
max }\sim 1/\gamma_\Vert$, we come to the system
\[\frac{{\rm d}I_\nu}{{\rm d}r}=-a_0I_\nu
I_{\nu_1},\]
\begin{equation}
\frac{{\rm d}I_{\nu_1}}{{\rm d}r}=a_0I_\nu I_{\nu_1},
\end{equation}
where
\begin{equation}
a_0\sim\frac{24n_er_e^2\gamma_0^2}{m\nu^2\theta^4\gamma^5},
\end{equation}
$\gamma$ and $\gamma_0$ stand for the characteristic values of
these quantities for a given particle distribution. It should be
noted that the system (13) along with equation (14) are literally
the same as those for the scattering off the rectilinearly moving
particles, except for the factor $\gamma_0^2$ entering equation
(14) \citep[cf. eqs. (8)-(9) in][]{p07a}. Generally speaking, one
can write that $\gamma_0^2/\gamma^5=1/\gamma_\Vert^5\gamma_0^3$
and conclude that the scattering by spiraling particles with the
longitudinal Lorentz-factor $\gamma_\Vert$ is $\gamma_0^3$ times
less efficient than that by the particles streaming along the
magnetic field with the same Lorentz-factor $\gamma_\Vert$.
However, in pulsar case the situation is somewhat different. As
has been shown in \citet{lp98}, the synchrotron absorption, which
determines the evolution of the particle distribution function,
acts to slow down the longitudinal motion and increase the
transverse momenta in such a way that
$\gamma=\gamma_0\gamma_\Vert$ keeps constant as long as the
particle pitch-angle is small enough, $\gamma_0/\gamma <\theta$.
Thus, it is reasonable to compare the scatterings by the original
and evolved distributions of particles, in which case $\gamma$ is
the same. Then the spiraling particles scatter $\gamma_0^2$ times
more efficiently than the streaming particles. Note also that in
the two processes the directions of predominant scattering of the
radio beam photons differ, $\theta_1^{\rm max}\sim
1/\gamma_\Vert=\gamma_0/\gamma$ and $\theta_1^{\rm max}\sim
1/\gamma$, respectively, implying different relations between the
interacting frequencies, $\nu_1\sim\nu\theta^2\gamma^2/\gamma_0^2$
and $\nu_1\sim\nu\theta^2\gamma^2$.

The system (13) has the first integral,
\begin{equation}
I\equiv I_\nu+I_{\nu_1}={\rm const},
\end{equation}
and the solution is written as
\[I_\nu=\frac{I/x}{1+1/x},\]
\begin{equation}
I_{\nu_1}=\frac{I}{1+1/x},
\end{equation}
where $x\equiv [I_{\nu_1}^{(0)}/I_\nu^{(0)}]\exp (\Gamma_0)$ and
$\Gamma_0\equiv Ia_0r$. The latter quantity characterizes the
scattering efficiency, whereas the former one the extent of
intensity transfer from the radio beam to the background. As long
as $x\ll 1$, $I_{\nu_1}\sim I_{\nu_1}^{(0)}\exp (\Gamma_0)$ and
$I_\nu\approx I_\nu^{(0)}$. At $x\sim 1$ $I_{\nu_1}$ becomes
comparable with $I_\nu^{(0)}$ and enters the stage of saturation.
Given that $x\gg 1$, $I_{\nu_1}\approx I_\nu^{(0)}$ and $I_\nu\sim
I_\nu^{(0)}/x$. Since initially the intensity ratio of the
background and the beam is extremely small,
$I_{\nu_1}^{(0)}/I_\nu^{(0)}<<<1 $, $\Gamma_0\sim n\times 10$ is
necessary to provide $x\sim 1$, in which case a substantial part
of the beam intensity is transferred to the background. As has
been shown in \citet{p07a}, in case of the scattering by the
streaming particles this condition may well be satisfied in
pulsars. For the scattering by the spiraling particles $\Gamma_0$
is $\gamma_0^2$ larger and, correspondingly, the growth of the
scattered component should be even more significant.

\subsection{The case $\omega^{(\rm c)}\sim s\omega_H/\gamma_0$,
$\omega_1^{(\rm c)}\ll \omega_H/\gamma_0$} Now we turn to induced
scattering between such two states that one of the frequencies is
close to the resonance, $\omega\gamma\eta\sim s\omega_H$, and
another one is well below the resonance,
$\omega_1\gamma\eta_1\equiv\omega\gamma\eta
-s\omega_H\ll\omega_H$. The intensity evolution in the two states
is given by
\[\frac{{\rm d}I_\nu^i}{{\rm d}r}=I_\nu^iI_{\nu_1}^j\frac{n_ec}{2}
\int\frac{\eta^2\nu}{\eta_1^3\gamma_\Vert^2\nu_1^4}\left
(\frac{{\rm d}\sigma_{\nu\to\nu_1}^{ij}}{{\rm d}O_1}\right )^{(\rm
c )}\frac{s\omega_H}{2\pi}
\]
\[\times\left [\frac{\partial f}{\partial p_\perp}\frac{mc}{p_\perp}+\frac{\partial f}
{\partial p_\Vert}\frac{mc\cos\theta}{p_\Vert\eta}\right ]p_\perp
{\rm d}p_\perp{\rm d}p_\Vert,
\]
\[
\frac{{\rm d}I_{\nu_1}^j}{{\rm
d}r}=-I_\nu^iI_{\nu_1}^j\frac{n_ec}{2}
\int\frac{\eta_1^2\nu_1}{\eta^3\gamma_\Vert^2\nu^4}\left
(\frac{{\rm d}\sigma_{\nu_1\to\nu}^{ji}}{{\rm d}O}\right )^{(\rm c
)}\frac{s\omega_H}{2\pi}
\]
\begin{equation}\times\left [\frac{\partial f}{\partial
p_\perp}\frac{mc}{p_\perp}+\frac{\partial f} {\partial
p_\Vert}\frac{mc\cos\theta}{p_\Vert\eta}\right ]p_\perp {\rm
d}p_\perp{\rm d}p_\Vert,
\end{equation}
where the subscripts $i,j$ denote the polarization states of the
initial and final photons, $({\rm d}\sigma_{\nu\to\nu_1}^{ij}/{\rm
d}O_1)^{(\rm c )}$ is the cross-section of the scattering from
$\nu$ to $\nu_1$ in the guiding centre frame (it is given by eq.
(7) and should be expressed via the quantities of the laboratory
system), the components of the momentum increment are given by
equations (8) and (9) and it is taken into account that
$\omega_1\gamma\eta_1\ll\omega_H,\omega\gamma\eta$.

The symmetry of the scattering probability (4) with respect to
direct and inverse scatterings implies that $({\rm
d}\sigma_{\nu_1\to\nu}^{ji}/{\rm d}O)^{(\rm c )}=(\nu^4/\nu_1^4)
({\rm d}\sigma_{\nu\to\nu_1}^{ij}/{\rm d}O_1)^{(\rm c )}$, and one
can see that $\vert{\rm d}I_\nu^i/{\rm d}r\vert=\alpha\vert{\rm
d}I_{\nu_1}^j/{\rm d}r\vert$, where
$\alpha\equiv\nu\eta/\nu_1\eta_1\gg 1$. Comparing the increments
of the transverse and longitudinal momenta we find that $\Delta
p_\perp\partial f/\partial
p_\perp\sim(\theta^2p_\Vert^2/p_\perp^2)\Delta p_\Vert\partial
f/\partial p_\Vert$. In our consideration
$p_\perp/p_\Vert\ll\theta$, i.e. the particle pitch-angle is much
less than the photon propagation angle, and hence the contribution
of the longitudinal increment can be neglected. Note that
$p_\perp$ enters equation (17) via $f$ as well as via the argument
$s\xi$ of the Bessel function and its derivative in the
cross-sections (7),
$\xi\equiv\beta_0\sin\theta^{(c)}=p_\perp\sin\theta/p_\Vert\eta\sim
2p_\perp/p_\Vert\theta\ll 1$. One can see that integration of the
kinetic equations (17) over $p_\perp$ by parts results in a change
of the sign of the right-hand sides. Thus, in the course of the
scattering the photons are mainly transferred from high harmonics,
$\nu\sim s\omega_H/2\pi\gamma\eta$, to the zeroth one,
$\nu_1\ll\omega_H/2\pi\gamma\eta_1$.

The harmonic number $s$ enters ${\rm d}I_\nu/{\rm d}r$ via
$s^2J_s^2(s\xi)$ or $s^2J_S^{\prime^2}(s\xi)$, which peak at high
harmonics, $s^{\rm max }\sim\gamma_0^3$. Note, however, that
inside the light cylinder $s^{\rm max}$ corresponds to the
frequencies lying in the optical or soft X-ray range \citep[see,
e.g.,][]{p07c}, and the intensity of pulsar radiation in these
bands is much weaker than the radio intensity. Therefore the
induced scattering from high harmonics, $s\sim s^{\rm max}$, is
expected to be inefficient. In the present consideration, we leave
aside this problem and concentrate on the induced scattering of
the radio emission. As the magnetic field strength decreases with
distance from the neutron star, a given radio frequency passes
through the resonances of increasingly higher order. At the same
time, the number density of the scattering particles and the radio
intensity strongly decrease with distance, so that the scattering
becomes much weaker.

Thus, we are interested in the radio beam scattering at several
first harmonics of the gyrofrequency, which takes place within the
light cylinder. Then $s\xi$ is still a small quantity and one can
take approximately that
\begin{equation}
J_s(s\xi)\approx\frac{1}{s!}\left (\frac{s\xi}{2}\right )^s,\quad
J_s^{\prime}(s\xi)\approx\frac{1}{2(s-1)!}\left
(\frac{s\xi}{2}\right )^{s-1}.
\end{equation}
Performing integration in equation (17), taking into account that
$\beta_0\approx 1$, $\theta\ll 1$, $\cos\theta^{(\rm
c)}=(\cos\theta -\beta_\Vert)/\eta\approx -1$, $\Delta\phi^{(\rm c
)}=\Delta\phi$ and noticing that $\theta_1^{\rm
max}\sim\gamma_0/\gamma$, one can come to the system
\[\frac{{\rm d}I_\nu^i}{{\rm d}r}=-\alpha a_s^{ij}I_\nu^iI_{\nu_1}^j,\]
\begin{equation}
\frac{{\rm d}I_{\nu_1}^j}{{\rm d}r}=a_s^{ij}I_\nu^iI_{\nu_1}^j,
\end{equation}
with
\[a_s^{AA}=a_s^{BA}=\frac{r_e^2n_e}{8m\gamma^3\nu_1^2\theta^2}
\left (\frac{\omega_H/2\pi\gamma}{\nu_1\eta_1}\right
)\frac{s^{2s+2}}{(s!)^2}\left (\frac{\gamma_0}{\gamma\theta}\right
)^{2s-6},\]
\[a_s^{AB}=a_s^{BB}\]
\begin{equation}
=\frac{r_e^2n_e}{64m\gamma^3\nu_1^2\theta^2}
\left (\frac{\nu_1\eta_1}{\omega_H/2\pi\gamma}\right
)\frac{(4+2s)s^{2s+1}}{(s!)^2}\left
(\frac{\gamma_0}{\gamma\theta}\right )^{2s-6}.
\end{equation}
One can see that $a_s^{AA}/a_s^{AB}=a_s^{BA}/a_s^{BB}\sim
(\omega_H/2\pi\gamma\nu_1\eta_1)^2\gg 1$, i.e. the scattering into
the state with A-polarization is much more intense. Hence, the
scattered component should be dominated by A-polarization,
similarly to the scattering between the frequencies below the
resonance (see Sect. 3.1). However, in contrast to that case, the
initial intensities in the two polarizations are affected
identically.

The general forms of the systems (13) and (19) differ by the
presence of the factor $\alpha$ in the latter one. Taking notice
that the substitution $\widetilde{I}_{\nu_1}=\alpha I_{\nu_1}$
makes equation (19) similar to equation (13), we immediately write
the first integral,
\begin{equation}
I\equiv I_\nu +\alpha I_{\nu_1}={\rm const},
\end{equation}
and the solution
\begin{equation}
I_\nu=\frac{I/y}{1+1/y},\quad I_{\nu_1}=\frac{I/\alpha}{1+1/y},
\end{equation}
where $y\equiv[\alpha I_{\nu_1}^{(0)}/I_\nu^{(0)}]\exp (\Gamma_s)$
and $\Gamma_s\equiv Ia_sr$. One can see that the intensity
transfer becomes significant at $y\ga 1$. Although the qualitative
character of the solution (22) is the same as that in case of the
scattering below the resonance, the quantitative difference is
substantial. The maximum possible intensity of the scattered
component appears much less, $I_{\nu_1}\sim I_\nu^{(0)}/\alpha\ll
I_\nu^{(0)}$, the rest of the energy being deposited to the
scattering particles. Note that the scattered intensity may still
be comparable with the original radio beam intensity at the same
frequency $\nu_1$. Indeed, the conditions
$\nu_1\ll\nu\eta/\eta_1\sim\nu\theta^2\gamma^2/\gamma_0^2$ and
$\theta^2\gamma^2/\gamma_0^2\gg 1$ allow  that $\nu_1\gg \nu$.
Then, with the decreasing spectrum of the radio beam, even a small
part of the beam intensity $I_\nu(\nu)$, which is transferred to
the background, may be strong enough as compared to
$I_\nu(\nu_1)$.

\section{Application to the Crab pulsar}
Now we turn to numerical estimates of the scattering efficiency
and start from the scattering between the frequencies below the
resonance. The growth of the scattered component at the frequency
$\nu_1$ is characterized by the quantity
$\Gamma_0=I_\nu^{(0)}a_0r$, where it is taken that $I\equiv
I_\nu^{(0)}+I_{\nu_1}^{(0)}\approx I_\nu^{(0)}$ and $a_0$ is given
by equation (14) with $\nu=2\nu_1/\theta^2\gamma_\Vert^2$. The
spectral intensity of the pulsar radio beam can be presented as
\begin{equation}
I_\nu^{(0)}=I_{\nu_0}\left (\frac{\nu}{\nu_0}\right )^{-\alpha}.
\end{equation}
Here $\nu_0\sim 10^8$ Hz and
\begin{equation}
I_{\nu_0}=\frac{L}{\nu_0S},
\end{equation}
where $L$ is the total radio luminosity of the pulsar, $S=\pi
r^2w^2/4$ is the cross-section of the pulsar beam at a distance
$r$ and $w$ is the pulse width in the angular measure. The number
density of the scattering particles can be written in terms of the
multiplicity factor of the plasma, $\kappa$,
\begin{equation}
n_e=\frac{\kappa B}{Pce},
\end{equation}
where $P$ is the pulsar period. With the dipolar geometry of the
magnetic field, $B\propto r^{-3}$, the scattering efficiency can
be estimated as
\[\Gamma_0=25\frac{0.1\,{\rm s}}{P}\frac{L}
{10^{30}\,{\rm erg\,s}^{-1}}\frac{B_\star}{10^{12}\,{\rm G}}
\frac{\kappa}{10^2}\left (\frac{10^9\,{\rm Hz}}{\nu_1}\right )^2
\left (\frac{\nu}{10^8\,{\rm Hz}}\right )^{-\alpha}\]
\begin{equation}
\times\left (\frac{w}{0.4}\right )^{-2}\left (\frac{r}{10^8\,{\rm
cm }}\right )^{-4}\left (\frac{\gamma}{10^2}\right )^{-5}\left
(\frac{\gamma_\Vert}{10}\right )^4\gamma_0^2,
\end{equation}
where $B_\star$ is the magnetic field strength at the surface of
the neutron star and the radius of the star is taken to be $10^6$
cm. The scattered component grows significantly provided that
$x\equiv [I_{\nu_1}^{(0)}/I_\nu^{(0)}]\exp (\Gamma_0)\sim 1$. As
has been shown in \citet{p07a}, this condition is satisfied for
$\Gamma_0\sim$ 20--30. From equation (26) above one can conclude
that such values of $\Gamma_0$ can indeed be characteristic of the
Crab pulsar.

It should be noted that the pulsar radio beam is broadband and its
angle of incidence increases with distance from the neutron star,
$\theta\propto r$. Therefore at different altitudes $r$ the
background component of a given frequency
$\nu_1=\nu\theta^2(r)\gamma_\Vert^2/2$ is fed by the radiation of
different frequencies $\nu\propto r^{-2}$. The number density of
the scattering particles and the incident intensity entering
$\Gamma_0$ are known to decrease with distance, $n_e\propto
r^{-3}$ and $I_\nu^{(0)}\propto r^{-2}$. However, at larger
altitudes the feeding frequency $\nu$ is lower, $I_\nu^{(0)}$ is
much larger, and for steep enough spectrum of the radio beam the
scattering is more efficient. In case of the Crab pulsar
$\alpha\approx 3$, so that $\Gamma_0\propto r^2$ and in the course
of the scattering the intensity is transferred to the background
from the lowest frequencies, the process taking place at distances
of order of the cyclotron resonance radius.

The efficiency of the scattering from several first harmonics of
the gyrofrequency, $\Gamma_s=I_\nu^{(0)}a_sr$, can be estimated
analogously. Substituting equations (23)-(25) into the first line
of equation (20), we obtain
\[\Gamma_s=0.5\frac{s^{2s+2}}{(s!)^2}\frac{\omega_H/2\pi}
{\nu_1\eta_1\gamma}\left (\frac{\gamma_0}{\gamma\theta}\right
)^{2s-4}\frac{0.1\,{\rm s}}{P}\frac{L} {10^{30}\,{\rm
erg\,s}^{-1}}\]
\[\times\frac{B_\star}{10^{12}\,{\rm G}}
\frac{\kappa}{10^2}\left (\frac{10^9\,{\rm Hz}}{\nu_1}\right )^2
\left (\frac{\nu}{10^8\,{\rm Hz}}\right )^{-\alpha}\]
\begin{equation}
\times\left (\frac{w}{0.4}\right )^{-2}\left (\frac{r}{10^8\,{\rm
cm }}\right )^{-4}\left (\frac{\gamma}{10^2}\right )^{-5}\left
(\frac{\gamma_\Vert}{10}\right )^4\gamma_0^2.
\end{equation}
The scattering of the lowest frequencies of the pulsar radio beam
is most efficient and it occurs just beyond the corresponding
resonance, at the distance satisfying the condition
$\nu\theta^2\gamma/2=s\omega_H/2\pi+\nu_1\gamma/\gamma_\Vert^2$
($\nu_1\gamma/\gamma_\Vert^2\ll\nu\theta^2\gamma/2,\,s\omega_H/2\pi$).

Comparison of equations (26) and (27) shows that
\begin{equation}
\frac{\Gamma_s}{\Gamma_0}\sim\frac{\nu_s\eta}
{\nu_{1_s}\eta_1}\left (\frac{\gamma_0}{\gamma\theta}\right
)^{2s-4}\left (\frac{\nu_0}{\nu_s}\right )^\alpha\left
(\frac{\nu_{1_0}}{\nu_{1_s}}\right )^2,
\end{equation}
where $\nu_s$ and $\nu_{1_s}$ are the frequencies involved in the
scattering at the harmonics of the gyrofrequency, whereas $\nu_0$
and $\nu_{1_0}$ participate in the scattering below the resonance.
Taking into account the relations $\nu_0\eta=\nu_{1_0}\eta_1$ and
$\nu_s\eta\gg\nu_{1_s}\eta_1$, one can see that for the scattering
of the same frequency, $\nu_s=\nu_0$,
$\Gamma_s/\Gamma_0\sim(\nu_s\eta/\nu_{1_s}\eta_1)^3
(\gamma_0/\gamma\theta)^{2s-4}$. As $\gamma_0/\gamma\theta\ll 1$,
we conclude that the scattering from several first harmonics of
the gyrofrequency (at least for $s=1,2$) may dominate the
scattering below the resonance. It should be kept in mind,
however, that these scatterings take place at different altitudes
in the magnetosphere, below and above the resonance, and because
of the spatial dependence of $n_e$ and $I_\nu^{(0)}$ the
scattering at the harmonics of the gyrofrequency is somewhat less
efficient.

The original radio beam is known to be directed along the magnetic
field in the emission region. Because of the magnetosphere
rotation, at large enough altitudes $r$ it makes the angle $\sim
r/2r_L$ with the local magnetic field direction (here $r_L=5\times
10^9P$ cm is the light cylinder radius). In the scattering regimes
considered in the present paper, the radio beam photons are
predominantly scattered along the ambient magnetic field in the
frame corotating with the neutron star. The rotational aberration
in the scattering region shifts the wavevector of the scattered
radiation by $\sim r/r_L$ in the direction of the magnetosphere
rotation, so that in the laboratory frame it makes the angle $\sim
r/2r_L$ with the original radio beam. Correspondingly, in the
pulse profile the scattered component precedes the MP in pulse
longitude by $\Delta\lambda\sim r/2r_L$, where $r$ is the
characteristic altitude of the scattering region \citep[for more
details see ][]{p07a}. As the radio beam scattering at the
harmonics of the gyrofrequency occurs at larger altitudes in the
magnetosphere as compared to the scattering below the resonance,
it should result in the components with larger separations from
the MP.

To have a notion about the component location in the pulse profile
let us estimate the radius of cyclotron resonance, $r_c$.
Proceeding from the resonance condition
$\nu\gamma\theta^2/2=\omega_H/2\pi$ and taking into account that
$\omega_H\propto B\propto r^{-3}$ and $\theta\approx r/2r_L$, we
find
\begin{equation}
\frac{r_c}{r_L}=\left (\frac{0.1\,{\rm s}}{P}\right )^{3/5} \left
(\frac{B_\star}{10^{12}\,{\rm G}}\frac{10^9\,{\rm Hz}}{\nu}
\frac{10^2}{\gamma}\right )^{1/5}.
\end{equation}
One can see that in the Crab pulsar the region of cyclotron
resonance of the radio frequencies lies close to the light
cylinder radius. Although for low enough frequencies the resonance
is slightly beyond $r_L$, we assume that the estimate
$\Delta\lambda\sim r/2r_L$ is still roughly applicable. The
location of the LFC $\sim 30^\circ$ ahead of the MP implies that
the scattering region is close to the light cylinder. As the PR
precedes the MP by $\sim 15^\circ$, it should originate in the
outer magnetosphere. Thus, we conclude that the PR component
results from the scattering below the resonance, whereas the LFC
arises in the course of the first-harmonic scattering.

\section{Discussion}
We have examined the process of magnetized induced scattering off
spiraling particles. Our consideration is restricted to the
scattering of radio frequencies in the magnetic fields well below
the critical value. In contrast to the classical problem on the
scattering by rectilinearly moving particles, in our case the
scattering at the harmonics of the particle gyrofrequency is not
negligible. In the course of spontaneous scattering off the
spiraling particles, the waves of the frequencies well below the
first resonance are mainly scattered to high harmonics, into the
range of the spectral maximum of the particle synchrotron emission
\citep{p07c}. The induced scattering, however, appears most
efficient for the pairs of states corresponding to close
harmonics. We have concentrated at the low-frequency scattering,
in which case one of the frequencies is well below the resonance
and another one corresponds to one of the several first harmonics.
It has been shown that the scattering in these regimes may
dominate the scattering between the pair of frequencies below the
resonance. The latter process is analogous to the scattering off
the rectilinearly moving particles, but it appears $\sim
\gamma_0^2$ times more efficient provided that the scattering
particles have the same total Lorentz-factor $\gamma$.

In application to pulsars, we are interested in the induced
scattering of a narrow radio beam into the background. Given that
the scattering is efficient, the photons are mainly scattered
along the ambient magnetic field, in the direction corresponding
to the maximum scattering probability, and the scattered component
may grow roughly as large as the original radio beam. In the
course of the scattering, the intensity is transferred from the
harmonics of the gyrofrequency into the state below the resonance.
The numerical estimates show that the scattering from several
first harmonics and between the frequencies below the resonance
can be substantial, especially for low enough frequencies of the
radio beam, in which case the incident intensity is the largest
because of the decreasing spectrum of the beam.

The scattered radiation makes the angle $\sim r/2r_L$ with the
incident radio beam (where $r$ stands for the altitude of the
scattering region) and precedes the MP in pulse longitude. The
scattering from different harmonics takes place at different
altitudes in the magnetosphere and therefore results in different
components in the pulse profile. The scattering between the
frequencies below the resonance has the lowest characteristic
altitude and gives rise to the component closest to the MP. The
components formed as a result of the scattering from increasingly
high harmonics should have increasingly large separations from the
MP. In application to the Crab pulsar, one can identify the
component resulting from the scattering below the resonance with
the PR and the component resulting from the first-harmonic
scattering with LFC. The MP-LFC separation, $\Delta\lambda\sim
30^\circ$, implies that the first-harmonic scattering occurs close
to the light cylinder. This is consistent with the estimate of the
location of the resonance region in this pulsar.

The formation of the LFC close to the light cylinder is strongly
supported by the polarization data: The position angle of linear
polarization in the LFC is shifted by $\sim 30^\circ$ from that of
the MP. Note that the position angle of the MP radiation is
determined by the magnetic field direction in the emission region,
whereas the position angle of the LFC radiation should reflect the
magnetic field orientation in the scattering region. Because of
the magnetosphere rotation, the ray emitted along the magnetic
field makes the angle $\sim r/2r_L$ with the local magnetic field
direction \citep[see][and Sect. 4 above]{p07a}. Thus, the magnetic
field orientations in the scattering and emission regions differ
by $\sim r/2r_L$, the difference between the position angles of
the original and scattered radiation being approximately the same.
For the scattering taking place close to the light cylinder this
difference is $\sim 30^\circ$. It is worthy to point out that in
our consideration the position angle shift of the scattered
component roughly equals its longitudinal separation from the MP.
This is indeed the case for the LFC of the Crab pulsar
\citep{mh98} and is believed to be a distinctive feature of the
scattered components in other pulsars.

High percentage of linear polarization is another characteristic
feature of the scattered components. In case of the scattering in
a strong magnetic field, the scattered radiation is dominated by
the waves of ordinary polarization, whose electric vector lies in
the plane of the wavevector and the external magnetic field. In
the original radio beam, only the ordinary waves are subject to
the scattering below the resonance, whereas both types of waves,
the ordinary and extraordinary ones, undergo equally efficient
scattering at the harmonics of the gyrofrequency. In the Crab, as
well as in other pulsars, the PR component is known to have almost
complete linear polarization. Although the percentage of linear
polarization of the LFC is lower, $\sim 40\%$, it still exceeds
that of the MP.

One can expect that the LFC radiation suffers depolarization. This
can be understood as follows. A noticeable sweep of the position
angle across the LFC \citep{mh98} implies that this component is
formed by the radiation coming from somewhat different altitudes
in the magnetosphere, which results from the scattering of
somewhat different frequencies. If one take into account the
finite width of the MP, at a fixed pulse longitude within the LFC
there should be radiation from different altitudes and therefore
with different position angles. The superposition of the waves
with different position angles may actually lead to a substantial
depolarization of the resultant radiation.

The LFC and PR of the Crab pulsar are known to exhibit pronounced
frequency evolution \citep{mh96}. The PR component is significant
at the lowest frequencies, the LFC becomes strong at frequencies
$\sim 1$ GHz, and at higher frequencies both components vanish. To
analyze the spectral behaviour of the scattered components in our
model let us turn to equation (28) and consider the ratio of the
scattering efficiencies given that the frequencies of the
scattered radiation are equal, $\nu_{1_0}=\nu_{1_s}$. Then we have
$\Gamma_s/\Gamma_0\sim(\nu_s\eta/\nu_{1_s}\eta_1)^{1-\alpha}
(\gamma_0/\theta\gamma)^{2s-4}$, i.e. the role of the scattering
at the harmonics of the gyrofrequency increases with frequency,
$\Gamma_s/\Gamma_0\propto\nu_{1_s}^{\alpha-1}$. Thus, the LFC is
expected to dominate at somewhat higher frequencies, which is in
accordance with the observed trend.

On the way in the magnetosphere, both scattered components, the PR
and LFC, may be further subject to scattering. Because of
magnetosphere rotation their inclination to the ambient magnetic
field rapidly increases with distance, while the magnetic field
strength rapidly decreases. Similarly to the MP, the components
may be involved in the induced transverse scattering in a
moderately strong magnetic field, in which case the radiation is
scattered backwards \citep[for the general theory of this process
see][]{p07b}. The consequences of the backward scattering of the
components will be studied in detail in a separate paper. It will
be shown that this process may give rise to the high-frequency
components in the profile of the Crab pulsar: the backscattering
of the PR can account for the IP', whereas the scattering of the
LFC can explain the HFC1 and HFC2. It will also be demonstrated
that the efficiency of the induced transverse scattering increases
with frequency and, correspondingly, at high enough frequencies
the PR and LFC vanish, their intensities being almost completely
transferred to the backward components.

The scattering efficiencies given by equations (26) and (27)
depend on the intensity of the incident radio beam and on the
number density and the characteristic Lorentz-factor of the
scattering plasma particles. All these quantities may show marked
pulse-to-pulse fluctuations, so that the scattering efficiencies
may vary as well. If the scattering is so strong that the
component growth is at the stage of saturation, $x\gg 1$ or $y\gg
1$, the fluctuations of $\Gamma$ do not affect the intensity of
the scattered component significantly. On condition that $x\sim 1$
or $y\sim 1$, however, even small fluctuations of the scattering
efficiency may lead to drastic variations of the scattered
component. According to equations (26)-(27), this condition may be
satisfied in a number of pulsars, which have large enough radio
luminosities, strong magnetic fields and short periods. Such
pulsars are believed to exhibit occasional activity at the pulse
longitudes preceding the MP. Namely, these pulsars are expected to
show the transient components with the spectral and polarization
properties similar to those known for the PR and LFC of the Crab
pulsar. Furthermore, the transient components resulting from the
higher-harmonic scattering can also be present in pulsar profiles,
in particular, in the Crab pulsar.

\section{Conclusions}
The components of pulsar profiles outside of the MP are known to
exhibit a number of peculiar properties, and at the same time the
pulse-to-pulse fluctuations generally testify to a physical
relation of these components to the MP. We believe that the
components outside of the MP originate as a result of induced
scattering of the pulsar radio beam into the background, with
different types of the components corresponding to different
scattering regimes. In the present paper, we have considered the
magnetized induced scattering off the spiraling particles, which
may be present in the outer magnetosphere of a pulsar. In this
case the scattering at the harmonics of the particle gyrofrequency
may be efficient. Our investigation is aimed at explaining, at
least partially, the extremely complex radio emission pattern of
the Crab pulsar. It has been demonstrated that the scattering from
the first harmonic of the gyrofrequency into the state below the
resonance can account for the formation of the LFC, whereas the
scattering between the states below the resonance can explain the
origin of the PR component.

As the scattered radiation concentrates along the local magnetic
field direction in the scattering region, the scattering in
different regimes, which hold at different altitudes, does form
different components in the pulse profile because of the
magnetosphere rotation. The observed LFC separation from the MP,
$\Delta\lambda\sim 30^\circ$, implies that the first-harmonic
scattering takes place close to the light cylinder. It is
important to note that the formation of the LFC at large enough
altitude in the magnetosphere and its orientation along the
ambient magnetic field are strongly supported by the observed
shift of the position angle of linear polarization with respect to
that of the MP. The position angle shift appears approximately
equal to the LFC-MP separation in pulse longitude. This is also
expected from an analysis of the ray-magnetic field geometry in
the rotating magnetosphere.

In the two regimes considered in the present paper, the scattering
mainly results in the waves of the ordinary polarization, whereas
both the ordinary and extraordinary waves are believed to be
present in the original radio beam. Hence, the PR and LFC should
be strongly polarized, which is consistent with the observational
data.

As long as the scattering efficiency is large enough for the
growth of the scattered component to reach the stage of
saturation, the frequency dependence of $\Gamma$ and the
variations of $\Gamma$ due to the pulse-to-pulse fluctuations of
the incident radio beam intensity and of the parameters of the
scattering plasma do not affect the scattered component
substantially. This is believed to be the case for the PR and LFC
of the Crab pulsar. However, the variations of the scattered
component can be dramatic provided that $\Gamma$ is somewhat
smaller and the scattering is about to reach the saturation stage.
Therefore we conclude that a number of pulsars may have the
transient components, which precede the MP and are analogous to
the PR and LFC in the Crab pulsar. The presence of the transient
components resulting from the higher-harmonic scattering is not
excluded as well.




\begin{thebibliography}{55}

\bibitem[\protect\citeauthoryear{Blandford \& Scharlemann}{1976}]{bs76}
Blandford R. D., Scharlemann E. T., 1976, MNRAS, 174, 59

\bibitem[\protect\citeauthoryear{Cordes et al.}{2004}]{h05}
Cordes J. M., Bhat N. D. R., Hankins T. H., McLaughlin M. A., Kern
J., 2004, ApJ, 612, 375

\bibitem[\protect\citeauthoryear{Dyks, Zhang \& Gil}{Dyks et al.}{2005}]{d05}
Dyks J., Zhang B., Gil J., 2005, ApJ, 626, L45


\bibitem[\protect\citeauthoryear{Fowler, Wright \& Morris}{Fowler et al.}{1981}]{f81}
Fowler L. A., Wright G. A. E., Morris D., 1981, A\&A, 93, 54

\bibitem[\protect\citeauthoryear{Fussell, Luo \&
  Melrose}{Fussell et~al.}{2003}]{melr_b}
Fussell D., Luo Q., Melrose D. B., 2003, MNRAS, 343, 124

\bibitem[\protect\citeauthoryear{Gil et al.}{1994}]{g94}
Gil J. A. et al., 1994, A\&A, 282, 45

\bibitem[\protect\citeauthoryear{Hankins \& Eilek}{2007}]{hc07}
Hankins T. H., Eilek J. A., 2007, ApJ, 670, 693

\bibitem[\protect\citeauthoryear{Jessner et al.}{2005}]{slow05}
Jessner A., Slowikowska A., Klein B., Lesch H., Jaroschek C. H.,
Kanbach G., Hankins T. H., 2005, AdSpR, 35, 1166

\bibitem[\protect\citeauthoryear{Kramer et al.}{1998}]{k99}
Kramer M., Xilouris K. M., Lorimer D. R., Doroshenko O., Jessner
A., Wielebinski R., Wolszczan A., Camilo F., 1998, ApJ, 501, 270

\bibitem[\protect\citeauthoryear{Luo \& Melrose}{2001}]{melr_a}
Luo Q., Melrose D. B., 2001, MNRAS, 325, 187

\bibitem[\protect\citeauthoryear{Lyubarskii \& Petrova}{1996}]{lp96}
Lyubarskii Yu. E., Petrova S. A., 1996, Astron. Lett., 22, 399

\bibitem[\protect\citeauthoryear{Lyubarskii \& Petrova}{1998}]{lp98}
Lyubarskii Yu. E., Petrova S. A., 1998, A\&A, 337, 433

\bibitem[\protect\citeauthoryear{Manchester \& Johnston}{1995}]{mj95}
Manchester R. N., Johnston S., 1995, ApJ, 441, L65

\bibitem[\protect\citeauthoryear{Manchester, Huguenin \& Taylor}{Manchester et al.}{1972}]{mht72}
Manchester R. N., Huguenin G. R., Taylor J. H., 1972, ApJ, 174,
L19

\bibitem[\protect\citeauthoryear{McCulloch et al.}{1976}]{m76}
McCulloch P. M., Hamilton P. A., Ables J. G., Komesaroff M. M.,
1976, MNRAS, 175, 71P

\bibitem[\protect\citeauthoryear{Moffett \& Hankins}{1996}]{mh96}
Moffett D. A., Hankins T. H., 1996, ApJ, 468, 779

\bibitem[\protect\citeauthoryear{Moffett \& Hankins}{1999}]{mh98}
Moffett D. A., Hankins T. H., 1999, ApJ, 522, 1046

\bibitem[\protect\citeauthoryear{Petrova}{2002}]{p02}
Petrova S. A., 2002, MNRAS, 336, 774

\bibitem[\protect\citeauthoryear{Petrova}{2003}]{p03}
Petrova S. A., 2003, MNRAS, 340, 1229

\bibitem[\protect\citeauthoryear{Petrova}{2004a}]{p04a}
Petrova S. A., 2004a, A\&A, 417, L29

\bibitem[\protect\citeauthoryear{Petrova}{2004b}]{p04b}
Petrova S. A., 2004b, A\&A, 424, 227

\bibitem[\protect\citeauthoryear{Petrova}{2008a}]{p07a}
Petrova S. A., 2008a, MNRAS, in press
(doi:10.1111/j.1745-3933.2007.00401.x)

\bibitem[\protect\citeauthoryear{Petrova}{2008b}]{p07b}
Petrova S. A., 2008b, ApJ, 673, 400

\bibitem[\protect\citeauthoryear{Petrova}{2008c}]{p07c}
Petrova S. A., 2008c, MNRAS, in press
(doi:10.1111/j.1365-2966.2007.12594.x)

\bibitem[\protect\citeauthoryear{Rankin et al.}{1970}]{r70}
Rankin J. M., Comella J. M., Craft H. D., Richards D. W., Campbell
D. B., Counselman C. C., 1970, ApJ, 162, 707

\bibitem[\protect\citeauthoryear{Vandenberg et al.}{1973}]{v73}
Vandenberg N. R., Clark T. A., Erickson W. C., Resch G. M.,
Broderick J. J., Payne R. R., Knowles S. H., Youmans A. B., 1973,
ApJ, 180, L27

\bibitem[\protect\citeauthoryear{Weltevrede, Wright \& Stappers}{Weltevrede et al.}{2007}]{welt07}
Weltevrede P., Wright G. A. E., Stappers B. W., 2007, A\&A, 467,
1163

\bibitem[\protect\citeauthoryear{Wilson \& Rees}{1978}]{wr78}
Wilson D. B., Rees M. J., 1978, MNRAS, 185, 297


\end{thebibliography}

\bsp

\label{lastpage}

\end{document}